\documentclass[lefttitle,11pt]{elsarticle}

\usepackage{amsmath}

\usepackage{fullpage}
\usepackage{subfigure}
\usepackage{enumitem}
\usepackage{hyperref}
\usepackage{xspace}
\usepackage{microtype}

\newcommand{\deltaPatch}{\ensuremath{\delta_c}}
\newcommand{\ie}{\emph{ie.}\xspace}
\newcommand{\eg}{\emph{e.g.}\xspace}
\begin{document}

\begin{frontmatter}
\title{Will you infect me with your opinion?}

%\reviewcopy{true} 

\author[1]{Krzysztof Domino}
\author[1]{Jaroslaw Adam Miszczak}
%\affiliation[1]{organization={Institute of Theoretical and Applied Informatics, Polish Academy
%		of Sciences},
%	addressline={Baltycka 5},
%	postcode={44-100},
%	city={Gliwcie},
%	country={Poland}}
%\affil[1]{Institute of Theoretical and Applied Informatics, Polish Academy of Sciences, Ba{\l}tycka 5, 44-100 Gliwice, Poland}
\ead{jmiszczak@iitis.pl}

%\author{K. Domino \and J.A. Miszczak}

%\usepackage{natbib}
%\setcitestyle{authoryear,round,aysep={}}

\begin{abstract}
Opinion formation is one of the most fascinating phenomena observed in human
communities, and the ability to predict and to control the dynamics of this
process is interesting from the theoretical as well as practical point of view.
Although there are many sophisticated models of opinion formation, they often
lack the connection with real life data, and there are still sociological
processes that need to be explained. To address this, we propose a model
describing the dynamics of opinion formation which mimics the process of the
virus or disease spreading in the population. The introduced model is motivated
by the model of disease spread with three possible channels -- direct contact,
indirect contact, and contact with ``contaminated'' elements. We demonstrate
that the presence of ``contaminated'' elements, which in the case of on-line
communities can be represented as the content published on the Internet, has
considerable impact on the process of opinion formation. We argue that by using
a simple mechanism of opinion spreading via passive elements, the introduced
model captures the meaningful elements of opinion formation in complex
communities. The presented work provides a step towards formulating universal
laws governing social as well as physical or technical systems.
\end{abstract}

%\tableofcontents

%%%%%%%%%%%%%%%%%%%%%%%%%%%%%%%%%%%%%%%%%%%%%%%%%%%%%%%%%%%%%%%%%%%%%%%%%%%%%%%%
%\section*{Keywords}
%%%%%%%%%%%%%%%%%%%%%%%%%%%%%%%%%%%%%%%%%%%%%%%%%%%%%%%%%%%%%%%%%%%%%%%%%%%%%%%%
\begin{keyword}
opinion transfer \sep disease spreading \sep agent interactions \sep direct and indirect communication \sep SIS model \sep oscillation of opinion intensity
\end{keyword}

\end{frontmatter}

%\maketitle

%%%%%%%%%%%%%%%%%%%%%%%%%%%%%%%%%%%%%%%%%%%%%%%%%%%%%%%%%%%%%%%%%%%%%%%%%%%%%%%%
\section{Introduction}
%%%%%%%%%%%%%%%%%%%%%%%%%%%%%%%%%%%%%%%%%%%%%%%%%%%%%%%%%%%%%%%%%%%%%%%%%%%%%%%%

The ability to predict and to control the opinion formation in communities is of
great theoretical and practical importance~\citep{lippmann2017public}. The power
to influence the formation of opinion in the human society would have tremendous
impact on many fields of human activity, including finances, medicine,
criminology, and political science. Methods for shaping public opinion are
particularly important in the era of rapid development of social media and their
increasing impact on all areas of social
live~\citep{bovet2016validation,gargiulo2017role,zhuravskaya2020political,petrova2021social,lee2022effect},
including the political competition. The opinion formation is important for
economic issues such as the stock market, where the organization of agents -- in
this case representing investors -- around some opinion can be the cause of the
bubble (\ie a rapid increase) or the crash of the assets'
prices~\citep{gligor2001econophysics, sornette1998hierarchical}. This stock
market behaviour can also be modelled by the complex network, see
eg.~\citep{park2020link}. Such complex network model is also the case for other
social and social-related phenomena, including the Internet and information
flow~\citep[Chapter~4]{chen2014fundamentals}, and epidemic
spreading~\citep[Chapter 5]{chen2014fundamentals}. Mentioned analogue gives the
motivation for our research in light of application of physical model,
concerning random walk process and information diffusion in particular, for the
opinion formation modelling.

There are variety of applications of physical models to
socialphenomena~\citep{castellano2009statistical,sobkowicz2009modelling,edmonds2013simulating}. These ideas start from Ettore Majorana~\citep{di1942valore}, who emphasized a possible analogy between statistical physics and social sciences, and follow the famous book of John Maynard Keynes in 1934~\citep{keynes2018general}, who pointed out that variations of stock assets prices have their origin in the collective crowd behaviour of interacting agents. Sixteen years after Keynes, inadequacy and inefficiency of existing theories describing social or financial systems was pointed out, \eg by George Soros~\citep{soros2015alchemy}. For current review of models of financial market and social behaviour in analogy to various physical models one can refer to~\citep{kutner2019econophysics}. For the recent review of the current state of the art and the promising directions for future research in the use of physics methods to study social phenomena one is advised to consult~\cite{jusup2022social}.

One of our goals is to enrich the field of social physics by concentrating on the opinion formation process itself, and extending the analogy between this process and the epidemic spreading~\citep{daley1964epidemics}. We
aim to reconstruct the ``wave like'' dynamics (or chaotic fluctuations) that is
present in opinion quantity~\citep{acemouglu2013opinion}, political
conviction~\citep{kramer1971short}, stock market assets
prices~\citep{bak1997price}, and epidemic
outbreaks~\citep{herrera2011multiple}. Although fluctuations appears in some of
sociological models mentioned in this paragraph, the above mentioned ``wave like''
dynamics is not the basic finding there, as the researchers are rather interested in
the steady state. Therefore, such fluctuations are rather tied to some particular
features of the opinion/knowledge transfer process, \eg some language
misunderstanding~\citep{foerster2018finite}, or the particular impact of
stubborn agents~\citep{acemouglu2013opinion}. Our goal is to check whether these
fluctuations can be derived from a more general model.

One should note that opinion formation displays a dynamic similar to the
knowledge transfer, and both processes that are often network
based~\citep{reagans2003network}. In particular, Kowalska-Stycze\'n et
al.~\citep{kowalska-styczen2018model} considered a network of informal contacts
as a crucial element underlying of knowledge dissemination in a community and
provided a model based on cellular automata. Nguyen et al.~\citep{nguyen2021}
studied a model of cultural dissemination, with an added element of social
influence. The results demonstrated that the more one allows individual agents to be
influenced by other agents in their local environment, the less homogeneous the
resulting society becomes. In real, however, world there are many non-homogeneous
social groups. In~\citep{axelrod1997dissemination} the agent based model was
applied to explain the difference between individuals in believes, attitudes, and behaviour. The main transfer of agents' attitude was from the neighbourhood, and
it was shown that the local convergence can create the global polarization. The
analogical mechanism of ``neighbourhood tipping'' was applied to explain social
segregation~\citep{schelling1971dynamic} and the split of society into two
parts.

There are analogical models of stock exchange agents divided into two
groups~\citep{bak1997price}, rational  -- analysing the real value of companies -- and
irrational -- just following others -- ones~\citep{kruszewska2013method}. If the first group dominated in size, we
would have the steady period with some variability of share prices. If the
second group dominated, we would have high variability and potentially the
bubble and the crisis. This observation is also tied to some sort of the
organization of agents, \eg the Ising model
in~\citep{gligor2001econophysics}. This is in some analogy to
\citep{sznajd2000opinion} where a simple Ising model of opinion formation in
closed community was proposed. This model -- know as United We Stand, Divided We
Fall or Sznajd-Weron model -- provides a powerful tool for describing the
dynamics of the behaviour of agents under the influence of their community. It
concludes that the common decision can not been made in the ``democratic way''
by a closed community. Following such approach, in~\citep{sznajdweron2005who}
the Ising model was applied to simulate the political process of opinion
formation in a society where two Ising variables for each agent are investigated
-- one concerning the economical freedom and one concerning the personal freedom.
Based on this, the agent may be: Authoritarian, Conservative, Libertarian, or
Socialist, yielding ACLS model.

The main contribution of the presented paper is the study of the opinion
formation based on the model constructed in the analogy to disease spread in the
population. We focus on the agent-based model of human population. The class of
such models is of interest to both physics and control systems communities.
Following the classification in~\citep{mastroeni2019agent}, we propose the
discrete model in the opinion domain, and the agent is either ``infected'' by
some opinion or they are is clear of it. One should note that the elementary
process of infecting is not symmetric, since only an ``infected'' agent can infect
others with their opinion. The agents who are not ``infected'' are
neutral from the model point of view. We consider three possible ways
of opinion transfer between agents motivated by the spectrum of disease
spreading channels: in the first channel it is pairwise, close neighbourhood interaction
 -- conversation and direct contact, in the second channel it is the further
neighbourhood contacts -- gossip propagation or casual engagement, and in the third channel it is the environment
long-range ``contamination''. In particular, the third channel can be
interpreted as the form of long-term presence in the \emph{environment}, which, in
the case of the Internet, represents posts and comments on web forums, and in 
more traditional forms can be understood as leaflets and advertising material
presented in our daily surroundings. This third channel has not yet been fully
explored in prevailing research. In our model, agents are allowed to move freely by the random walk
parameterized by the mobility factor, their spatial localization determines the
transfer via the particular channel or channels.

To make the presented model self-contained we also consider the opinion decay,
\ie we assume that at each time step some agents can get ``cured'' with the given probability. The opinion decay was used in~\citep{kimura2012opinion} where the opinion formation was modelled by the closed neighbours voting, but also including the past neighbours' opinions. To account for the fact that the old
opinion is getting outdated, the linear, experiential or power-law decay
function can be applied. Such decay appears empirically in many empirical
models, and for a sound example refer to Google search index on COVID in recent
years~\citep{husain2020covid}. After the rapid rise in the interest in this
topic, it decays, and then reappears again after the new proton of information
-- the opinion propagates, then decays, and propagates again. Note that in the
Ising model, often used to model opinion formation
\citep{gligor2001econophysics,sznajd2000opinion}, the opinion decay is modelled
by the parameter-dependent noise.

The remaining part of this work is organized as follows. In Section~\ref{sec:model}, we introduce a model proposed for study of disease spread as well as opinion formation and information transfer in communities. In Section~\ref{sec:num-exp}, we provide numerical results demonstrating main features of the introduced model. In Section~\ref{sec:discussion}, we discuss the strengths and the limitations of the presented model. Finally, in Section~\ref{sec:conclusions} we provide a summary of the presented results as well as some concluding remarks.

%%%%%%%%%%%%%%%%%%%%%%%%%%%%%%%%%%%%%%%%%%%%%%%%%%%%%%%%%%%%%%%%%%%%%%%%%%%%%%%%
\section{Three-channel model of opinion formation}\label{sec:model}
%%%%%%%%%%%%%%%%%%%%%%%%%%%%%%%%%%%%%%%%%%%%%%%%%%%%%%%%%%%%%%%%%%%%%%%%%%%%%%%%

In this work we consider a simple model of opinion spread based on the assumption that the opinion (or knowledge) can be transferred via one of three channels, namely:
\begin{enumerate}[label={(C\arabic*)}]
	\item \textbf{first channel} -- representing a direct interaction or direct contact with an agent propagating the opinion;
	\item \textbf{second channel} -- representing an indirect contact with an agent propagating the option; the channel is equivalent to information transfer from a close neighbourhood or via unintended interactions (eg. gossip or accidental information gain);
	\item \textbf{third channel} -- representing contact with an environment \emph{contaminated} by the agents propagating the opinion; the channel introduces the possibility of interaction between the agents via long-term memory.
\end{enumerate}

The first channel represents a direct contact between agents. This mechanism may represent a conversation or a directed engagement in a discussion in the case of information spreading. In this way, it is possible to provide full information about the disseminated facts. The channel captures the effect of the direct contact with an ``infected'' individual, in the case of infection spread. For each unit time step, if agents $i$ appears in the close neighbourhood of ``infected'' agent $j$, we can model the probability of agent $i$ getting ``infected'' by agent $j$, using notation from \cite{tabasso2019diffusion}, by $p^{c1}_{i,j} = \nu_1$,
where $0 \leq \nu_{c1} \leq 1$ is the rate at which information is transmitted at
the direct meeting i.e. the particular weight of the first channel. Bear in mind that in more detailed approach this rate can be variable accounting for the ineffectiveness of $j$ and the ``immunity'' of $i$. The first channel is most directly tied to the mobility of the agents, introduced later and denoted by $\mu$. In practice the mobility parameter can represent a physical mobility of a person or a activity of a person in the social network. Thus, the limitation in mobility will directly influence the effectiveness of this channel.

The second channel for dissemination of the opinion or knowledge is based on the indirect contact between the agents. This channel captures the mechanics underlying the gossip propagation or rumor spreading in communities~\citep{lind2007spreading, wang2017rumor, robbins2019who, banerjee2019using, zhou2021scir}. However, in the presented approach we are restricting the study to one opinion only. Hence, the gossip is understood not as \emph{false} information, but rather as the information which is obtained accidentally. Note that this mechanism is in analogy to the ``contamination'' of the place where agents are located (a room, compartment) \citep{vuorinen2020modelling}, where there is a possibility to get infected merely by spending time in the proximity of an ``infected'' agent. In the case of opinion spreading, we assume that for each ``infected'' agent in the surrounding area the probability of agent $i$ getting ``infected'' is 
$p^{c2}_{i} = \nu_2$,  
where $0 \leq \nu_2 \leq 1$ is the rate at which information is transmitted via environment, \ie the particular weight of the second channel. One should note that in more detailed approach one can also consider some factors tied to the size of the community. In the case of infection spreading, community contamination is proportional to the number of ``infected'' agents over some volume~\cite{vuorinen2020modelling}.

Finally, the third channel is based on the long-term contamination of some elements of the environment. The interpretation of this channel is clear in the case of disease spread where the infection can be caused by the contact with the contaminated material. A similar mechanism can be observed in the digital as well as physical environments. In the first case, the contaminated material represents the digital fingerprinting in the form of information posted in social networks, which can lead to the information (or disinformation) spread in the network. If ``infected'' agent appears in a certain location (patch), this patch can be ``infected'' with probability $\nu_{\text{patch}}$, called the \textit{patch contamination probability}. Then, if agent $i$ appears in the patch that has already been ``infected'', the probability of agent $i$ getting ``infected'' is
$p^{c3}_{i} = \nu_3$, where $0 \leq \nu_3 \leq 1$ is the rate at which information is transmitted from patches i.e. the particular weight of the third channel.

The ``infected'' patch is healed with the probability $\deltaPatch$ or stays ``infected'' with the probability $1-\deltaPatch$.
Additionally, in the analogy to infection spreading mechanism and SIR model~\citep{kermack1927contribution,harko2014exact}, we assume that each idea decays in time. At each time step if agent $i$ is ``infected'' (or ``informed'' or ``convinced''), it stays ``infected'' with probability $1-\delta$ or becomes ``cured'' with probability $\delta$. In the context of opinion formation, being cured means that the agent lost the interest in the promoted opinion or forgot about the disseminated information. However, in contrary to SIR model, we do not consider the death (removal) possibility for the agents as well as any form of immunity after ``infection''. Hence, our approach is in line with the SIS (susceptible-informed-susceptible) model. In such model agent can either be ``informed'' (or ``infected'') or susceptible.

We should also stress that all three channels considered in the described model can be interpreted in the physical as well as digital environment. In the first case, the direct communication is interpreted as face to face interaction, the second channel is interpreted as propagation via unintentional information gain, and the contaminated material is interpreted as a physical form of information distribution in the form of leaflets and posters. On the other hand, if we consider the propagation of information in the digital environment, the first channel is interpreted as direct discussion via email or instant messengers, the second channel can be understood as information obtained using mailing lists and mass mail distribution, and the third channel represents the content and comments published in the community. The main difference between the second and the third channel is that the last one has a potential for long-term impact on the opinion dynamics, and can be used to boost the opinion formation and stability.

%%%%%%%%%%%%%%%%%%%%%%%%%%%%%%%%%%%%%%%%%%%%%%%%%%%%%%%%%%%%%%%%%%%%%%%%%%%%%%%%
\section{Numerical study}\label{sec:num-exp}
%%%%%%%%%%%%%%%%%%%%%%%%%%%%%%%%%%%%%%%%%%%%%%%%%%%%%%%%%%%%%%%%%%%%%%%%%%%%%%%%

To investigate the role of of the presented model we developer a series of numerical simulations. Each numerical experiment is composed of simulation steps, and during every step, each individual (agent) in the population is able to move and can be infected vai one of the channels. 
In each simulation step we have the following probabilities of getting ``infected'':
\begin{enumerate}[label={(C\arabic*)}]
	\item with rate $\nu_1$ from other agents via direct contact, if $i$ is in direct contact with ``infected'' $j$;
	\item with rate $\nu_2$ from other agents via indirect contact, from each ``infected'' $j$ that in the surrounding of $i$;
	\item and with rate $\nu_3$ from the environment, if $i$ is on ``infected'' patch.
\end{enumerate}

One should note that the third channel is also parametrized by the path contamination probability $\nu_{\text{patch}}$, which controls the transfer of the opinion from ``infected'' agents to patches (environement).
Additionally, we assume simplistic mode of healing both for agents and patches.
The probability of an agent getting ``cured'' at each step is $\delta$, while the probability of a patch getting cured at each step is $\deltaPatch$.
Concluding, we have six free parameters related to the opinion transmission $\delta$, $\deltaPatch$, $\nu_1$, $\nu_2$, $\nu_3$, and $\nu_{\text{patch}}$, one free parameter $\mu$ controlling the mobility of agents.

%%%%%%%%%%%%%%%%%%%%%%%%%%%%%%%%%%%%%%%%%%%%%%%%%%%%%%%%%%%%%%%%%%%%%%%%%%%%%%%%
\subsection{Implementation details}
%%%%%%%%%%%%%%%%%%%%%%%%%%%%%%%%%%%%%%%%%%%%%%%%%%%%%%%%%%%%%%%%%%%%%%%%%%%%%%%%

To study the behaviour of our model, we provide a model implemented using NetLogo agent-based modelling environment~\citep{tisue2004netlogo}. For the sake of reproducibility~\citep{turing-way,Leipzig2021}, the source code for the presented numerical experiments -- including the model and the files with the specification of parameters used in simulation -- can be obtained from publicly accessible repository~\citep{three-way-repo}. Additionally, the model has been also published in the NetLogo Modeling Commons~\citep{v3w-mc}, where it can be run directly in the web browser. The parameters for controlling the model are divided into three sections. 

The global properties of the environment are fixed during the initial setup of the model. At this step one can fix the population (using slider \texttt{population}), and percentage of patches which are considered as unavailable for agents (slider \texttt{init-obstacles-ratio}). Moreover, it is possible to control the initial percentage of contaminated patches (using slider \texttt{init-contamination-ratio}) and the initial number of ``infected'' agents (slider \texttt{init-infected-ratio}).

The second group of properties is used to set the probabilities of the spread channels and includes three parameters of the opinion (or infection) spread. This group can be used to control the strengths of the three channels. For this purpose sliders specifying \texttt{direct-infection-prob}, \texttt{indirect-infection-prob}, and \texttt{patch-infection-prob} can be used to change the probabilities of getting ``infected'' via direct contact or first channel, $\nu_1$, indirect contact or second channel, $\nu_2$,, and via contact with a contaminated patch or third channel, $\nu_3$. The values of these probabilities are set independently.

The third group of parameters is related to the control pf the behaviour and properties of agents and patches. This includes the parameters controlling the healing and the mobility. In particular, one can control the following parameters of the model:
\begin{itemize}
\item $\mu$ -- mobility, understood as the probability that the agent will make a move at the simulation step, parameter controlled by the slider \texttt{mobility-prob};
\item $\delta$ -- probability that the ``infected'' agent will recover during the simulation step, parameter controlled by the slider \texttt{agent-healing-prob};
\item $\deltaPatch$ -- probability that the contaminated patch will be cured during the single simulation step, parameter controlled by the slider \texttt{patch-heal-prob};
\item $\nu_{\text{patch}}$ -- probability that the patch will be contaminated if visited by the ``infected'' agent, parameter controlled by the slider \texttt{patch-contamination-prob}.

\end{itemize}

One should note that the presented model does not include any dependency of the mobility of the agent on the agent's location. The mobility of the agent is constant over time. Thus, the mobility influences only the probability of changing the position from the current to one of available positions. Moreover, the mobility is constant for all agents. However, in the model used in the numerical simulation an additional parameter describing obstacles is included, which could be used to differentiate the behaviour of agents depending on their location. 

%%%%%%%%%%%%%%%%%%%%%%%%%%%%%%%%%%%%%%%%%%%%%%%%%%%%%%%%%%%%%%%%%%%%%%%%%%%%%%%%
\subsection{Results}
%%%%%%%%%%%%%%%%%%%%%%%%%%%%%%%%%%%%%%%%%%%%%%%%%%%%%%%%%%%%%%%%%%%%%%%%%%%%%%%%

The main novelty of the model introduced in this work is the presence of the additional, long-term channel supporting the propagation of the opinion (or the infection) without the interaction between the agents. To the best of our knowledge, this approach has not been studied in the recent literature devoted to the study of opinion formation using agent-based models~\citep{mastroeni2019agent}.

For this reason, in this study we are mostly interested in the long-range indirect interaction channel of the opinion propagation (third channel). Our main goal is to explore the relation between the opinion propagation via the contaminated media and the standard channels, \ie direct contact and the proximity contact, represented by first channel and second channel two in the model presented in Section~\ref{sec:model}.

As the third channel is the main focus of this paper and its role is to the large extend unexplored, we would like to argue that the appropriate utilization of this channel can explain some non-trivial phenomena of the opinion spreading. In particular, we check if the impact of third channel can be replaced by the second channel that correspond to indirect contact, and if it is possible to grasp the relation between the third channel and two channels traditionally considered in the agent-based models of opinion dynamics. Additionally, we are interested in the connection between the mobility -- be that in the physical or digital reality -- and the impact of the contaminated environment on the sustainability and stability of the formed opinion.

Furthermore, we are interested in the stability of the opinion formation. Thus, we explore the interplay of the configurations of the opinion propagation channels leading to the stable and unstable states. To achieve this, we search for such parameters settings where some chaotic oscillations or ``wave like'' behaviour can be observed. This would be a non-trivial result for opinion spreading.

Finally, for the sake of clarity, we stress that we study the simplest case where only one opinion is considered, and we do not include competing opinions in our numerical studies. For this reason, from the purpose of out study the figure of merit is the fraction of agents in the population ``infected'' by the opinion.

%%%%%%%%%%%%%%%%%%%%%%%%%%%%%%%%%%%%%%%%%%%%%%%%%%%%%%%%%%%%%%%%%%%%%%%%%%%%%%%%
\subsubsection{Agent-based and environment-based channels}
%%%%%%%%%%%%%%%%%%%%%%%%%%%%%%%%%%%%%%%%%%%%%%%%%%%%%%%%%%%%%%%%%%%%%%%%%%%%%%%%

In the first series of numerical experiments our goal is to assess the importance of the third channel on the process of opinion formation in comparison to the first and the second channel. To achieve this we consider the impact of the third channel under the varied mobility of the agents, and for different values of path contamination probability. Thus, we can observe how its impact changes depending on the parameters describing agent-based and environment-based channels. This enables us to observe  the changes in opinion dissemination depending on the relative strength of the dissemination channels, by including a wide range of parameters describing the mobility and the probability of contaminating environment.

To clarify the role of the third channel, we consider two scenarios of parameter manipulation. In both cases we consider a natural situation where all three channels play an active role in the process of opinion formation, but with different strengths, described by parameters $\nu_1$, $\nu_2$, and $\nu_3$.

In the first scenario, we fix the impact of the first channel and the second channel, namely $\nu_1$ and $\nu_2$, and vary the impact of the third channel, namely $\nu_3$. Such setup enables us to distinguish between the role of agent-based channels (C1) and (C2), and patch-based channel (C3). In particular, on the basis of this distinction, one can see that the character of channels (C1) and (C2) is identical as both of them require a proximity between the agents. Thus, it is sufficient to explore the interplay between one of the agent-based channels and the channel (C3).

\begin{figure}[ht!]
	\centering
	\includegraphics{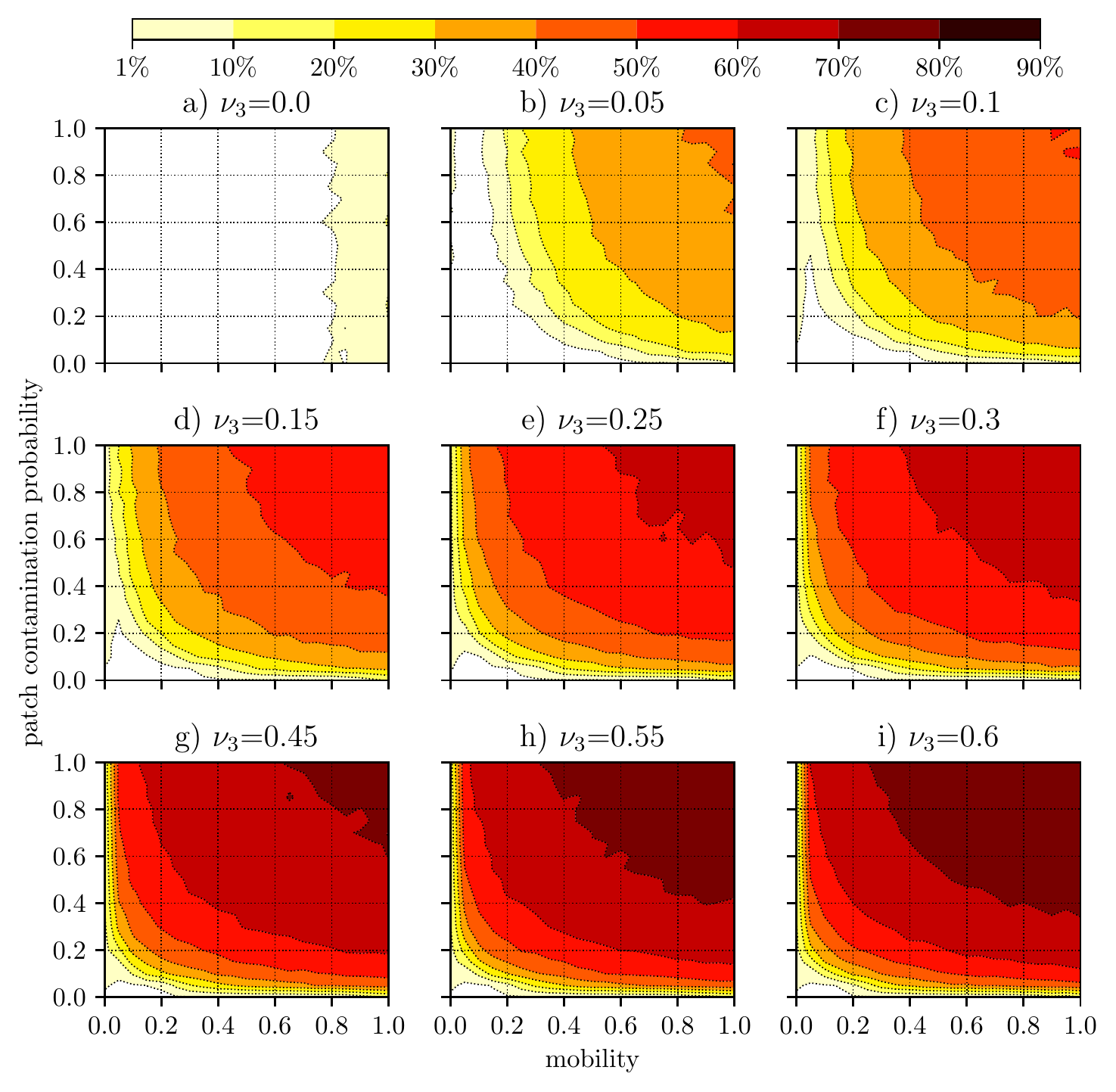}
	\caption{Dependency between the mobility-based and environment-based contamination channels. The plots illustrate the average percentage of ``infected'' agents for different values of weight $\nu_3$ assigned to  channel (C3) with fixed values of parameters $\nu_1$ and $\nu_2$. Each plot illustrates the values for patch contamination probability in range $\nu_{\text{patch}}=0,0.05,\dots,1$, and mobility in range $\mu=0,0.05,\dots,1$. Probabilities for getting infected via channels (C1) and (C2) are fixed to $\nu_1=0.5$ and $\nu_2=0.5$, respectively. Agent healing probability is $\delta=0.15$ and patch healing probability is $\deltaPatch=0.05$. The population is set to $250$ agents, the initial contamination of patches is $10\%$, the ``infected'' agents constitute initially $5\%$ of the population, and the obstacle ratio is $10\%$.  Average is calculated over $50$ realizations of the experiment, each consisting of $5000$ simulation steps. Note that in each experiment realization the location of mobility obstacles is randomly selected. One can observe that the increasing mobility can decrease the number of infected (convinced) agents for small values of mobility, suggesting that the transmission via contaminated patches provides a more stable mechanism for the opinion spreading.}
	\label{fig:mobility-patch-contamination-prob}
\end{figure}

The average number of infected (or convinced) agents in the scenario described above is presented in Fig.~\ref{fig:mobility-patch-contamination-prob}.
The plot in Fig.~\ref{fig:mobility-patch-contamination-prob}a) provides a sanity check, and one can see that for the value $\nu_3=0$, the propagation does not depend on the patch contamination probability.

On the other hand, for small values of $\nu_3$, as depicted in Figs.~\ref{fig:mobility-patch-contamination-prob}b) and \ref{fig:mobility-patch-contamination-prob}c), one can observe the interplay between spread due to channels depending on the mobility -- channels (C1) and (C2), and channel (C3). In particular, even for a small $\nu_3=0.05$, one can observe the noticeable number of infected agents, as long as the patch contamination probability is above $0.75$. This effect is observed for very low mobility. In fact, after increasing the mobility, the number of infected agents decreases. Thus, one can suggest that the mobility can reduce the effect of the third channel as long as the probability of getting infected by the environment is small. This effect can be most prominently observed in Figs.~\ref{fig:mobility-patch-contamination-prob}c) and d).

Furthermore, for the values of $\nu_3>0.25$, one can observe the effect of compensation of the small mobility by the infection through the channel (C3). For the values $\nu_3=0.25$ and $\nu_3=0.30$, as can be seen in Figs.~\ref{fig:mobility-patch-contamination-prob}e) and f), patch contamination probability, $\nu_\textrm{patch}$, and the mobility, $\mu$, can be seen as complementary factors when considering the average number of infected agents. However, for the values $\nu_3\geq0.45$, one can observe the asymmetry between $\nu_\textrm{patch}$ and $\mu$. In particular, if we consider values of $\nu_\textrm{patch}<0.2$ and $\mu\approx1$, presented in Fig.~\ref{fig:mobility-patch-contamination-prob}g), one can see that it is impossible to achieve the level of contamination which can be obtained by increasing the patch contamination, keeping the mobility significantly below $0.2$. This suggests that even for the case where the channels (C1) and (C2) have a larger probability of infecting agents, the action of channel (C3) leads to the increased number of infected agents. This effect also becomes visible in Figs.~\ref{fig:mobility-patch-contamination-prob}h) and i) for the higher values of $\mu$.

\begin{figure}[ht!]
	\centering
	\includegraphics{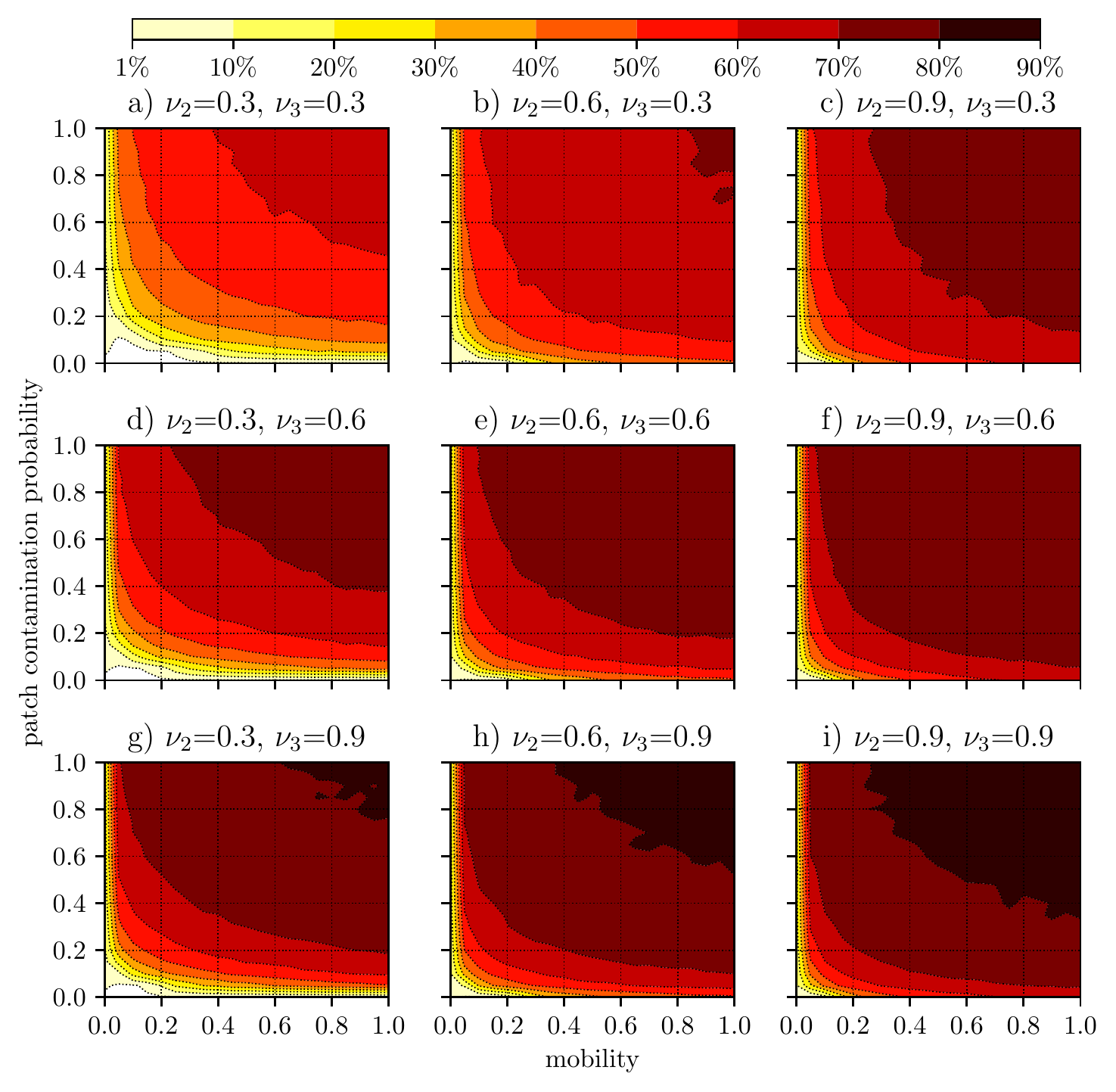}
	\caption{Dependency between the indirect infections (C2) and the infections via the  contaminated patches (C3). Each plot represented a percentage of ``infected'' agents for different values of $\nu_3$ and $\nu_2$ in the case of low $\nu_1 = 0.1$. Other parameters have the values according to the description in the caption of Fig.~\ref{fig:mobility-patch-contamination-prob}. In this experiment one can observer that  both indirect channels -- channel (C2) and channel (C3) -- have different impact on the opinion spread. In particular for $\nu_2=\nu_3=0.6$, it can be noticed that the increase in the mobility resulted in the faster increase of the number of infected agents. This suggest that mobility and channel (C2) is dominating for for $\nu_c\leq0.6$. However, one the most important effect is the much larger fraction of infected agents which can be obtained if we increase the value of $\nu_3$, even for small values of $\nu_2$. Thus, one can argue -- in particular panels c) and g) -- that the channel (C3) is crucial to achieve large number of infected (convinced) agents in the population.}
	\label{fig:mobility-patch-contamination-prob-two-channels}
\end{figure}

In the second scenario, where the simulation results are presented in~Fig.~\ref{fig:mobility-patch-contamination-prob-two-channels}, we consider a situation where the direct interaction based on the physical proximity has a fixed, small value. At the same time, we consider the changes in the number of the infected agents with the varying probabilities of getting infected via channels (C2) and (C3). Thanks to this, we can compare the separate impact of channels (C2) and (C3) on the process of opinion formation.
One should note that such scenario is easier to interpret in the case of the communication via digital media, than in the case of an infectious disease.
In the on-line communication, a direct interaction (eg. via chat) constitute only a fraction of interactions, and the two remaining channels are crucial for the opinion spread.

In the case presented in~Fig.~\ref{fig:mobility-patch-contamination-prob-two-channels}, we fix the value $\nu_1=0.1$, which makes the first channel significantly weaker than the other two channels.
In this setup, one can notice that the influence of both channel (C2) and (C3) is complementary, and one can utilize them interchangeably. As in the previous situation, the reduction in the number of infected agents due to the mobility is again visible only for small values of $\nu_2$ and $\nu_3$.

However, the most important observation is that by increasing the value of $\nu_3$ it is possible to achieve the infection ratio of over 80\% (Figs.~\ref{fig:mobility-patch-contamination-prob-two-channels}g) and h)), which was not possible when only the value of the indirect infection probability, $\nu_2$, was increasing (Figs.~\ref{fig:mobility-patch-contamination-prob-two-channels}c) and f)). This suggests that it is better to provide a robust means supporting channel (C3) instead of investing in channel (C2), which requires mobility of the agents. Additionally, in many cases, including the users in the digital environment, it is hard to control the mobility. Thus, opinion propagation via channel (C3)  provides means for controlling opinion dynamics which more robust, and, at the same time, easier to adjust.

As it was already mentioned, we are interested in the analysis of the interplay between the mobility of the agents and the role of  channel (C3). The mobility of agents can also be interpreted as their activity in the contaminating media. Thus, the mobility should have an impact on the effectiveness of channel (C3), by affecting the distribution of ``infected'' patches, as well as the average time agents spend on these patches. In this light, the important observation from Fig.~\ref{fig:mobility-patch-contamination-prob} is that for similar strength of all channels the plot is symmetric -- the impact of mobility is analogical as the impact of patch contamination probability. 

Furthermore, one can conclude, that channel (C3) plays a non-trivial role in information propagation, even if its strength is small.
See top panels of Fig.~\ref{fig:mobility-patch-contamination-prob}, where at small mobility levels the increase of
mobility would decrease information propagation. This is probably caused by the effect of mobility on the length of the stay of agents at infected patches. This gives some analogy to the phase transition behaviour, where the mobility would be the order parameter. Interestingly, such mobility can be tied to the``social temperature'', that is the noise factor introduced in Ising social model by~\citep{sznajdweron2005who}.

In Fig.~\ref{fig:mobility-patch-contamination-prob-two-channels} 
particular impacts of channels (C2) and (C3) are analysed, while the impact of channel (C1) is small and constant. The most important observation is that indirect channels are distinguishable, i.e. channel (C3) has different impact on the information spread than channel (C2) -- compare panels c) and g), and in this light the impact of channel (C3) is stronger. Furthermore, for high intensity of both (C2) and (C3) we have almost full information coverage -- see Fig.~\ref{fig:mobility-patch-contamination-prob-two-channels}i). As in case of Fig.~\ref{fig:mobility-patch-contamination-prob}, symmetry of particular plots suggests that the impact of mobility is analogical to the impact of patch contamination probability.

%%%%%%%%%%%%%%%%%%%%%%%%%%%%%%%%%%%%%%%%%%%%%%%%%%%%%%%%%%%%%%%%%%%%%%%%%%%%%%%%
\subsubsection{Stability of opinion formation}\label{sec:second_exp}
%%%%%%%%%%%%%%%%%%%%%%%%%%%%%%%%%%%%%%%%%%%%%%%%%%%%%%%%%%%%%%%%%%%%%%%%%%%%%%%%

Another aspect of the model proposed in this work is its effect on the stability of the opinion formation and transfer. As we are only interested in the number of agents ``infected'' during the process we will search for the fluctuation in the number of ``infected'' agents.

\begin{figure}[h!]
	\centering	
  \includegraphics{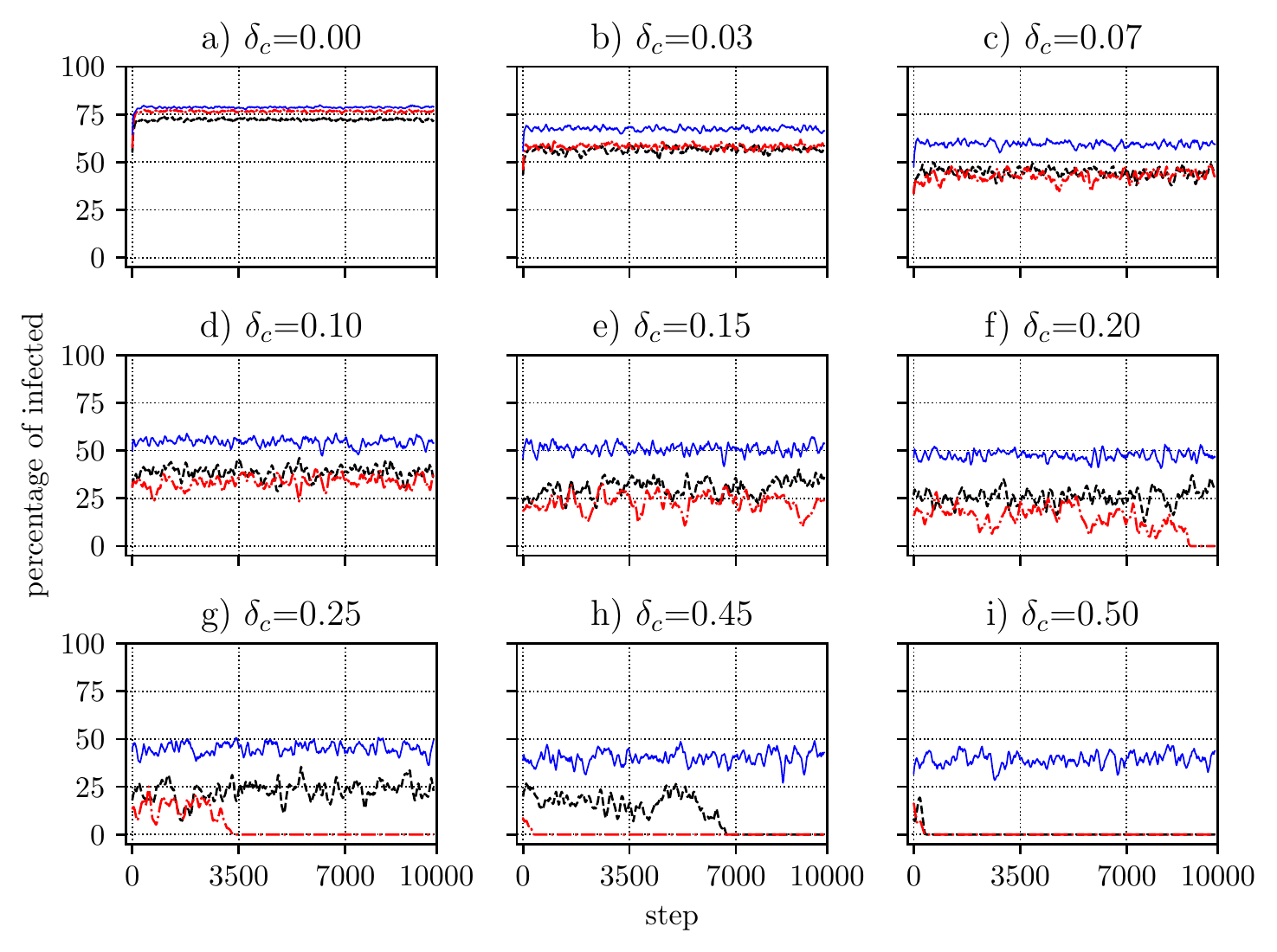}
  \caption{Impact of the patch healing on the stability of the opinion formation. Plots present single realizations of the opinion spread for different values of patch healing probabilities for situation with equal weights for all channels, $\nu_1=\nu_2=\nu_3=0.33$ (dashed black line) and in the situation with $\nu_1=\nu_2=0.25$ and $\nu_3=0.5$ (dash-dotted red line) $\nu_1=\nu_2=\nu_3=0.5$ (solid blue line). For the sake of clarity plots represent moving average of the percentage of ``infected'' agents for $100$ steps. Here $\nu_{\text{patch}}=0.15$, $\mu=0.5$ and $\delta=0.15$. Other parameters are the same as in Fig.~\ref{fig:mobility-patch-contamination-prob}. One should note that we can observe chaotic like behaviour for various parameter settings.   
Clearly, if the channel (C3) is dominating (dash-dotted red line), the opinion dynamics  suffers the most from the increasing probability of patch healing. This is equivalent to saying that to make channel (C3) relevant for the stable opinion formation, patch healing process has to be weak. What's more, in all cases the increase in the healing probability leads to a increasing instability of the number of infected agents. This provide another argument confirming the stabilizing effect of the contamination via infected patches. 
}
  \label{fig:patch-healing-prob-non-equal}
\end{figure}

The chaotic, or ``wave like'', behaviour is typical for some social phenomena, like those governing the financial market dynamics~\citep{gugler2012determinants}, or political voting behaviour~\citep{kramer1971short}. Other examples include fluctuation of U.S. public opinion on climate changes in 2002--2010~\citep{brulle2012shifting}, or variations of Google searches on COVID in recent years~\citep{husain2020covid}. In opinion formation research such fluctuations are modelled as rather special case caused by some particular phenomena such as non-perfect knowledge transfer due to some language misunderstanding~\citep{foerster2018finite}, or the presence of stubborn agents~\citep{acemouglu2013opinion}. 

Our goal is to reveal whether the model presented in this work can also account for similar fluctuations. For this purpose we study the examples of realizations of the opinion formation process. Again, we consider two scenarios: one with the all three channels contribution to the opinion formation, and one for investigating the dependency between the agent-based propagation and environment-based propagation.

\begin{figure}[ht!]
	\centering	
	\includegraphics{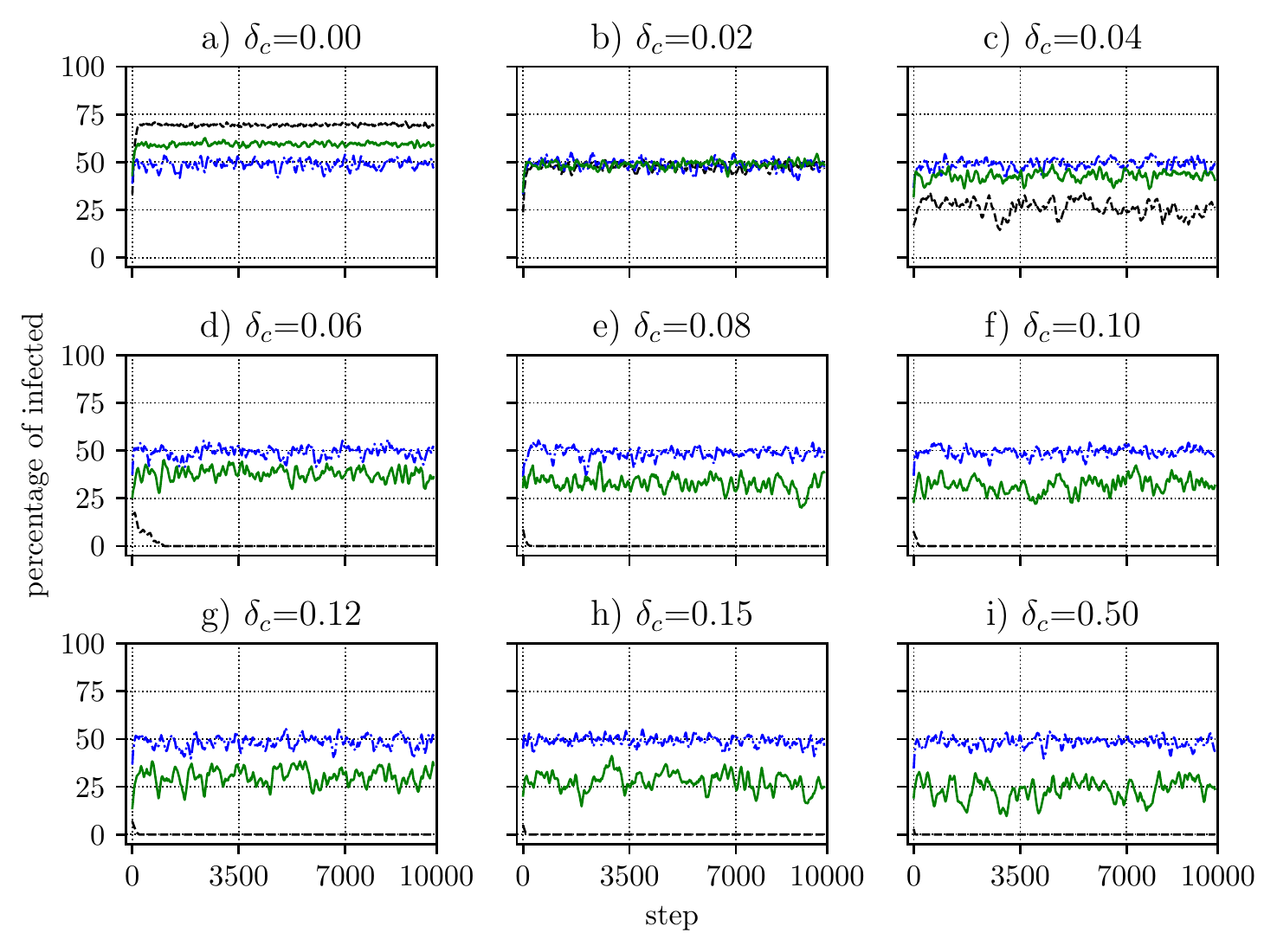}
	\caption{Impact of channel (C2) and the channel (C3) on the dynamics of the process. Some asymmetry between these channels can be observed, in a typical realization with fixed value of the direct contamination weight, $\nu_1=0.66$. Dashed black line represent situation with $\nu_2=0$, $\nu_3=0.33$, solid green line with $\nu_2=0.33$, $\nu_3=0.1$, dashed-dotted blue line represents situation $\nu_2=0.66, \nu_3=0$. Plots represent moving average of the percentage of ``infected'' agents for 100 steps. Agent healing ratio was set to $\delta=0.15$. Other parameters are the same as in the numerical experiments presented in Figs.~\ref{fig:mobility-patch-contamination-prob} and~\ref{fig:patch-healing-prob-non-equal}, \ie $\nu_{\text{patch}}=0.15$, $\mu=0.5$. Here, one can clearly see that even for small contribution of channel (C3) -- solid green line, $\nu_3=0.1$ -- the increase  in the patch healing results in the increased instability of the opinion formation. This, along with the observation that channel (C3) leads to a significant large number of infected (convinced) agents (cf. panel a)) supports the claim that opinion spread support with contaminated patches leads to opinion dynamics improved in terms of effectiveness and stability.
}
	\label{fig:patch-healing-prob-direct-vs-indirect}
\end{figure}

For the case with all three channels active during the opinion formation the resulting dynamics of the process is provided in Fig.~\ref{fig:patch-healing-prob-non-equal}. The figure includes time series of the intensity of an opinion, measured as the number of ``infected'' agents. 

First of all, one can observe that for patch healing probability $\deltaPatch=0$, provided in Fig.~\ref{fig:patch-healing-prob-non-equal}a), the number of infected agents is stable, and depends only on the probabilities of the channels. For the intermediate cases, $\deltaPatch=0.03$ to $\deltaPatch=0.15$, Figs.~\ref{fig:patch-healing-prob-non-equal}b)-e), it is clear that the patch healing has a destabilizing effect on the opinion formation. As expected, the case with the dominating channel (C3) suffers the most from the increasing probability of patch healing. For the values of $\deltaPatch\geq0.20$, the process of patch healing can lead to the situation where the population of infected agents dies off for high weight on channel (C3), $\nu_3$. Interestingly, for the case $\nu_1=\nu_2=\nu_3$, the effect is visible for $\deltaPatch=0.45$, which suggests that such high probability of patch healing, makes channel (C3) irrelevant for the stable opinion formation. 

Additionally, for some parameters settings, we have observed ``wave like'' chaotic fluctuations. For the values $\deltaPatch<0.10$ such fluctuations are not observed. Thus, as $\deltaPatch$ is the parameter of (C3) the provided examples suggest that channel can be used to stabilized the information propagation, or sustain the number of infected individuals.

In Fig.~\ref{fig:patch-healing-prob-direct-vs-indirect} we examine the situation where a fixed probability of getting infected via channel (C1) has a larger impact. Again, one can see that for $\deltaPatch=0$, the presence of channel (C3) has a stabilizing effect on the process of opinion formation,  Fig.~\ref{fig:patch-healing-prob-direct-vs-indirect}a). However, even a slight increase in patch healing probability leads to the destabilization and cancels out the advantage gain by using (C3), Fig.~\ref{fig:patch-healing-prob-direct-vs-indirect}b). Furthermore, in the scenario with dominating (C1), the population of infected agents dies off only in the situation where two other channels are unable to compensate the high values of $\deltaPatch$,
~\ref{fig:patch-healing-prob-direct-vs-indirect}d)-i). From this and form examples in Fig.~\ref{fig:patch-healing-prob-non-equal}f)-i), one can see that to compensate high probability of patch healing, other channels have to have high infection probability.

From the results presented in Figs.~\ref{fig:patch-healing-prob-non-equal} and \ref{fig:patch-healing-prob-direct-vs-indirect}, we can conclude that the effect of channel (C3) measured in terms of stability, depends on the so-called quality of this channel. This quality is measured by the patch healing -- or clearing -- probability, $\deltaPatch$. The lower the value of $\deltaPatch$ the higher the quality (or contagiousness) of the ``infected'' patches. Importantly, the large value of $\deltaPatch$, which can be understood as a presence of low-quality patches, leads to the less stable behaviour of the process. On the other hand, for small $\deltaPatch$ channel (C3) can boost the propagation significantly, see Fig.~\ref{fig:patch-healing-prob-non-equal}a). As such, as we have already seen in the previous subsection, the presence of channel (C3) can be beneficial for the purpose of maintaining high level of ``infected'' agents. For these reasons, in order to ensure a stable and high presence of opinion in the community, it is crucial to ensure that if the channel (C3) is present with high weight, the contaminated patches are cleared with low probability.

%%%%%%%%%%%%%%%%%%%%%%%%%%%%%%%%%%%%%%%%%%%%%%%%%%%%%%%%%%%%%%%%%%%%%%%%%%%%%%%%
\section{Discussion}\label{sec:discussion}
%%%%%%%%%%%%%%%%%%%%%%%%%%%%%%%%%%%%%%%%%%%%%%%%%%%%%%%%%%%%%%%%%%%%%%%%%%%%%%%%

In the presented numerical experiments we observed that the patch contamination probability value, $\nu_{\text{patch}}$, can be switched with the mobility value $\mu$, at least for the large weight of the third channel. This leads to the straightforward observation that patches can have a similar effect to the mobility, what is well justified if we use the analogy between patches and the Internet. Following this reasoning, we can consider the mobility of agents in light of the ``social temperature'', that is the noise factor introduced in social models by~\citep{sznajdweron2005who}. Hence, the Internet can also have an impact on this ``social temperature''. The impact of mobility coincides with an unusual observation. For small mobility, $0 \leq \mu \leq M0.1$, and  rather small impact of (C3), there is the sharp pick of information propagation versus mobility. This observation may suggest some form of the phase transition behaviour in the model.

As suggested in~\citep{sobkowicz2009modelling}, we link our findings to some real life data, to make the model more practical.
It has been observed that many opinion measures display the ``wave like'' oscillations, \eg in economics~\citep{gugler2012determinants}, politics~\citep{kramer1971short}, opinion on climate changes~\citep{brulle2012shifting}, and opinion on COVID~\citep{husain2020covid}. We have observed such ``wave like'' chaotic oscillations, and suggest that it is caused by the interplay between all channels. We have observed such dynamics in in the introduced mode for $\deltaPatch > 0$ in many cases. Most intense oscillations occurs, however, for large $\deltaPatch$ with (C3) active, $\nu_3=not=0$. This suggests some special role of this channel in the oscillation mechanism. One should keep in mind  that (C3) can be tied either with physical or Internet  medium. This coincides with the fact that oscillations in information spreading have already been examined in pre-Tnternet era~\citep{kramer1971short} where patches could rather be tied with handbills. On can suggest that the popularity of the Internet as a communication media leads to ubiquity of such chaotic patterns as there the $\deltaPatch$ parameter may be higher, which is related more frequent overwritten one information by another.

To asses the potential for further extension of our model, especially concerning the Internet, one should note that in the case of infection spread, there are important groups of super-spreaders~\citep{cave2020covid,small2006super,stein2011super}. Although we limit ourselves to one type of spreaders of opinion, it is important to mention that super-speeding has an analogy in opinion formation (as some particular bloggers, internet ``authorities'' etc.). The impact of super-spreaders of the opinion can be investigated as an elaboration of our model in the future.

One should also bear in mind that we have analyzed only one type of information, no matter whether it is correct or false. However, recently researchers refer to more types of information, some of them being fake. As an example we can refer to~\cite{merlino2020debunking} where SIS model with false information and debugging was applied. In such model, it was demonstrated that if the knowledge transfer rate is high enough (in comparison with the knowledge deterioration rate) the steady state of constant information amount in the network is achieved. Despite the fact that we deal only one type of information, the above mentioned findings comply (at least qualitatively) with our results, where rather constant non-zero fractions of ``infected'' agents were observed for highest knowledge transfer rates and lowest healing rate of patches. 

%%%%%%%%%%%%%%%%%%%%%%%%%%%%%%%%%%%%%%%%%%%%%%%%%%%%%%%%%%%%%%%%%%%%%%%%%%%%%%%%
\section{Conclusions}\label{sec:conclusions}
%%%%%%%%%%%%%%%%%%%%%%%%%%%%%%%%%%%%%%%%%%%%%%%%%%%%%%%%%%%%%%%%%%%%%%%%%%%%%%%

In this paper we proposed a model of infection propagation  based on three channels of propagation -- direct interaction, indirect interaction, and patch contamination -- and we utilized it to study the propagation of opinion in the community. Based on this model, we reconstructed many interesting social phenomena such as ``wave like'' oscillations on number of opinion caring agents, or the presence of ``infected'' patches that can mimic the role of the Internet. We  suggest that such chaotic patterns are caused, at least to some extent, by the the ``infected'' patches mechanism, and will be even more evident in future as the popularity of the Internet will grow. One can see the interaction based on the contaminated patches as the form of long-term memory, which in the Internet is present in the form of the content accumulated as the effect of the discussions or other interactions. This analogy is very interesting itself, as it is the step toward the unification of the description of some social and physical phenomena. As such, it follows the path set by Majorana, and then followed by Keynes, Soros and others.

The research is based on the author's hope that it is possible to formulate universal laws governing some social as well as physical or technical systems~\citep{goffman1964generalization,daley1964epidemics}. As it was suggest in \cite{helbing2014saving}, there are many situations, including crowd disasters, crime, terrorism, and the spread of disease, in which the careful study of the complex system can have a significant impact on saving endanger human lives. In particular, agent-based modelling provide a powerful tools which can support the formulation of laws hovering the complex system. Equipped with these tools, the model described in this work, even if motivated by some studies concerning the propagation of viral infections, touches today's world sociological problems. Nowadays, one can see many pseudo-scientific opinions often spread via the Internet, as virulent infection influencing the society. Such opinions can die off, hold on, or fluctuate depending on the parameters of susceptibility, mobility, and expiration. The presented work provides a step towards the understanding the conditions necessary to sustain the opinion, and could be helpful for estimating the cost of such action.

%%%%%%%%%%%%%%%%%%%%%%%%%%%%%%%%%%%%%%%%%%%%%%%%%%%%%%%%%%%%%%%%%%%%%%%%%%%%%%%
\subsection*{Acknowledgements}
%%%%%%%%%%%%%%%%%%%%%%%%%%%%%%%%%%%%%%%%%%%%%%%%%%%%%%%%%%%%%%%%%%%%%%%%%%%%%%%

The authors would like to thank Arkadiusz Sochan for interesting discussions concerning the simulation of virus propagation in indoors environment, and to Izabela Miszczak for proofreading the manuscript.

%%%%%%%%%%%%%%%%%%%%%%%%%%%%%%%%%%%%%%%%%%%%%%%%%%%%%%%%%%%%%%%%%%%%%%%%%%%%%%%
%\bibliographystyle{elsarticle-num}
%\bibliography{three-way}
%%%%%%%%%%%%%%%%%%%%%%%%%%%%%%%%%%%%%%%%%%%%%%%%%%%%%%%%%%%%%%%%%%%%%%%%%%%%%%%

\end{document}